\documentclass[prl, twocolumn,showpacs,superscriptaddress]{revtex4}
\usepackage{amsmath,amssymb}
\usepackage{graphicx}
\usepackage{pslatex}

\usepackage{color}
\newcommand{\rem}[1]{}

\begin{document}

\title{Spatial signal amplification in cell biology: a lattice-gas model for self-tuned phase ordering}
\author{T. Ferraro}
\email{teresa.ferraro@na.infn.it}
\affiliation{Dip. di Scienze Fisiche, 
Universit\`a di Napoli ``Federico II", via Cintia, 80126, Napoli, Italia}
\author{A. de Candia}
\affiliation{Dip. di Scienze Fisiche, 
Universit\`a di Napoli ``Federico II", via Cintia, 80126, Napoli, Italia}
\author{A. Gamba}
\affiliation{Politecnico di Torino and CNISM, Corso Duca degli Abruzzi 24, 10121 Torino, Italia}
\affiliation{INFN, via Pietro Giuria 1,  10125 Torino, Italia}
\author{A. Coniglio}
\affiliation{Dip. di Scienze Fisiche, 
Universit\`a di Napoli ``Federico II", via Cintia, 80126, Napoli, Italia}


\begin{abstract} 

Experiments show that the movement of eukaryotic cells is regulated by a
process of phase separation of two competing enzymes on the cell membrane,
that effectively amplifies shallow external gradients of
chemical attractant.
Notably, the cell is able to self-tune the final 
enzyme
concentrations to an equilibrium state of phase coexistence, for a wide
range of the average 
attractant
concentration. We propose a simple
lattice model in which, together with a short-range attraction between
enzymes, a long-range repulsion naturally arises from physical
considerations, that easily explains such observed behavior.

\end{abstract}

\maketitle

Specific moments of the life of a cell living in a multicellular
community, such as migration, proliferation, organization in layers or
complex tissues, imply spatial organization along some axis of
direction.
The original spatial symmetry of the cell must be broken to
adapt to a highly structured anisotropic environment.
For instance, migrating cells must orient towards sources of chemical
attractants, mitotic cells must orient along the spindle-pole axis to bud
daughter cells, epithelial cells must recognize the inner and outer part
of tissues to define organ boundaries.
From a physical point of view, these spatial organization phenomena may be
seen as self-organized phase ordering processes, where the cell state,
spontaneously, or because driven by an external field, decays into a state
of coexistence of two or more %
chemical phases, spatially
localized in different regions in order to define a front and a
rear, a top and a bottom, an outer and an inner part.
Local thermodynamic equilibrium requires precise tuning of
chemical potentials to nongeneric values to allow the coexistence of
different phases.
To implement this requirement in a robust way, capable of giving a stable
response over a wide range of stimulation situations, biological
systems must be endowed with
self-organized tuning mechanisms leading to phase coexistence
and polarization.

Directional sensing in chemotacting eukaryotic cells provides 
a beautiful
illustration of these principles~\footnote{The biological facts presented
below are taken, where not otherwise specified, from the
reviews~\cite{RSB+03} and~\cite{LH96}.}.
At the heart of directional sensing lies a chemical phase
separation process taking place on the inner surface of the cell membrane~\cite{GCT+05}.
The main players of the process are the enzymes 
phosphatidylinositol 3-kinase (PI3K), and 
phosphatase and tensin homolog (PTEN), 
which catalyze the switch of 
the phospholipid phosphatidylinositol between the bisphosphate (PIP$_2$)
and the trisphosphate (PIP$_3$) states.
The phospholipids are permanently bound to the inner
face of the cell membrane, while the two enzymes diffuse in the cell
volume and become active when they are adsorbed on the membrane.
PI3K adsorption takes place through binding to receptors of the external
attractant.
PTEN adsorption takes place through binding to the PTEN product, PIP$_2$,
a process which introduces an amplification loop in the system
dynamics~\cite{ID02,GCT+05}.  
A second amplification loop provided by PI3K binding to PIP$_3$~\cite{DMY+06}
has been recently observed.

Although there are no relevant
enzyme--enzyme or phospholipid--phospholipid interactions, the above
described catalytic processes, together with phospholipid diffusion on the
cell membrane, mediate an effective short-range attractive interaction
among enzymes of the same type.  
This interaction drives the system towards phase separation in a
PTEN-rich and a PI3K-rich phase ~\cite{GCT+05,CGC+07}. 
Two different regimes of
membrane polarization may be distinguished. 
In the presence of an attractant gradient,
anisotropy driven polarization
is realized in a time of
the order of a few minutes, and results in the formation of a PI3K-rich
patch on the membrane side closer to the attractant source and of a
PTEN-rich patch in the complementary region~\cite{ID02}. 
The process works as an efficient gradient amplifier:
a few percent gradient is sufficient to
completely polarize the cell membrane.
The orientation response is reversible:
by inverting the gradient direction the polarization orientation is also
inverted.
On the
other hand, cells exposed to uniform distributions of attractant
polarize in random directions over a longer timescale.
The average concentration of attractant is of crucial
importance, as shown by experimentally observed dose-response
curves~\cite{Zig77}: 
directional sensing does not take place neither at very low nor at very high
attractant levels, and there exists an optimal attractant concentration such
that the cell response is maximal.

On the basis of a simple analogy with 
the physics of binary mixtures, one would expect that the coexistence 
between the PI3K-rich and the PTEN-rich phase
would require a fine tuning
of the chemical potential difference between the two species.
Surprisingly, phase separation takes place instead for a wide range of absolute
concentration of the attractant, and therefore of absolute values of the
chemical potential for PI3K adsorption.
To explain this mechanism we propose here a simple lattice-gas model
in which, together with the effective short-range attraction
between enzymes, a long-range repulsion naturally
arises from the finiteness of the enzymatic reservoir, 
that easily explains the observed
behavior.

\textit{Model --} We represent the cell membrane by a square lattice of
size \(L\) with \(N\) sites, using periodic boundary conditions.
The sites $i$ occupied by PI3K (PTEN) are described by a $S_i=+1$ ($-1$) spin
~\footnote{One can imagine 
performing a coarse-graining of the system on an appropriate length scale
and associating a $+1$ sign to PI3K-rich sites
and a $-1$ sign to PTEN-rich sites.}.
We denote by \( N^\pm_\mathrm{tot} \) the total
number of \( \pm 1 \) enzymes in the cell, which is given by
the sum of the number of cytosolic (free) enzymes and the
number of membrane-bound enzymes:
$N^\pm _ \mathrm{tot}=N^\pm _ \mathrm{free}+N^\pm$.
The probability that a PI3K enzyme binds to site $i$ is proportional to
the number of cytosolic PI3Ks and to 
the density of binding sites (activated receptors with local concentration 
$c_i$ and PIP$_3$'s).
As a first approximation, the PIP$_3$ concentration can be
assumed to be
linearly dependent from the
density of PI3Ks.
This gives, on site $i$:
\begin{equation}
{\cal P}(-1\to +1)\propto \left[c_i+
\alpha^+\left(c^+_0+\beta^+\sum\limits_{j\in\partial i}S_j\right)\right]
N^{+}_\text{free}
\end{equation}
where $\alpha^+,\,\beta^+,\,c_0^+$ are 
functions of the chemical reaction rates, and $\partial i$ 
are the nearest neighbors of $i$.
Similarly, the probability that a PTEN 
molecule binds to site $i$ is proportional
to the number of free PTENs, and to 
the concentration of PIP$_2$:
\begin{equation}
{\cal P}(+1\to -1)\propto \alpha^-\left(c^-_0-\beta^-\sum\limits_{j\in\partial i}S_j\right)
N^{-}_\text{free}
\end{equation}
We interpret $\Delta{\cal H}=\ln[{\cal P}(-1\to +1)/{\cal P}(+1\to -1)]$
as an energy difference (in units of $k_BT$) between states $S_i=+1$ and $S_i=-1$, 
depending
both on
the local field $\sum_{j\in\partial i}S_j$ and on the number of cytosolic PI3Ks and PTENs.
Since
$N^{+}+N^{-}=N$, we can express $N^{+}_\text{free}$ and $N^{-}_\text{free}$ 
as functions
of the magnetization $m=(N^{+}-N^{-})/N$. 
Linearizing 
$\Delta{\cal H}$ 
around $\sum_{j\in\partial i}S_j=0$ and $m=0$, we obtain:
\begin{equation}
\Delta{\cal H}=-2J\sum\limits_{j\in\partial i}S_j-2h_i+2\lambda\, m
\label{deltaH}
\end{equation}
where $J=\frac{1}{2}\left(\frac{\alpha^+\beta^+}{c_i+\alpha^+ c_0^+}+\frac{\beta^-}{c_0^-}\right)$,
$h_i=\frac{1}{2}\ln\left(1+\frac{c_i}{\alpha^+ c_0^+}\right)-h_0$, with $h_0=
\frac{1}{2}\ln\left(\frac{\alpha^-c_0^-m^+}{\alpha^+ c_0^+m^-}\right)$, and
$\lambda=\frac{1}{2}\left(\frac{1}{m^+}+\frac{1}{m^-}\right)$, with
$m^\pm=2N^\pm_\text{tot}/N-1$. If 
$\frac{\beta^+}{c_0^+}<\frac{\beta^-}{c_0^-}$ 
we can neglect the dependence of $J$ on the attractant 
concentration $c_i$.

Eq.\ (\ref{deltaH}) corresponds to the variation of the Hamiltonian
\begin{equation}
{\cal H}=-J\sum_{\langle ij\rangle}S_iS_j - \sum_i h_iS_i +\frac{\lambda}{N}\sum_{i<j}S_iS_j.
\label{hamilton}
\end{equation}
The model 
(\ref{hamilton}) contains
a short-range ferromagnetic interaction representing the effective
attractive interaction between enzymes,
a long-range antiferromagnetic interaction 
which results from the finiteness of the
cytosolic enzymatic reservoir,
and an external site-dependent field
representing the effect of the attractant.
The latter depends on the concentration $c_i$ of activated receptors, which we take
proportional to the 
concentration of external attractant, in the form
$c_i=c(1+\epsilon\sin^2{\frac{\pi x_i}{L}} \sin{\frac{2\pi y_i }{L}})$.

When $h_i$ is
independent of $i$, the second and third term of Eq.\ (\ref{hamilton})
can be written (apart from a constant) as $\frac{N\lambda}{2}(\frac{h}{\lambda}-m)^2$,
so that energy minimization leads the system to self-tune to the
magnetization value $h/\lambda$.
Eq. (\ref{deltaH}) shows that $S_i$ is subject to the action of an
effective external field $h_{\mathrm{eff},i}=h_i-\lambda\,m$. 
The value $h_{\mathrm{eff},i}$ measures the degree of metastability of the
PTEN phase, and tends to zero during the self-tuning evolution of the
system. 
To realize the phase separation, $J$ has to be greater than the
critical value for the two-dimensional Ising spin model
($J\simeq 0.44$).
Furthermore, in the absence of chemical attractant,
the membrane has to be fully occupied by PTEN, 
implying $h_0\ge 1$.
We set for definiteness 
$J=h_0=\lambda=\alpha^+ c_0^+=1$.

\begin{figure}[htb]
\includegraphics[height=54mm]{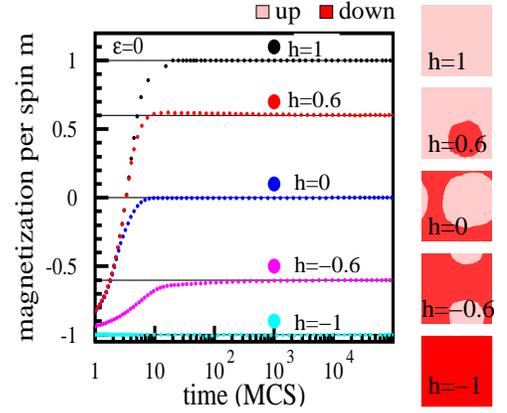}
\vspace{-0.3cm}
\caption{
Self-tuning dynamics in the presence of a uniform activation field $h$.
The magnetization $m$ grows to compensate the external activation field $h$. 
On the right, equilibrium states corresponding to different values of $h$.
}
\label{mag}
\end{figure}

\textit{Simulations --}
We study by Monte Carlo simulations
the dynamics and the final state attained by the system,
using a square lattice of size $L=2048$.
We first consider the case $\epsilon=0$, which corresponds to uniform stimulation.
In the absence of stimulation ($c=0$, implying $h=-1$ and $m=-1$)
the membrane is uniformly populated by PTEN molecules.
Setting $c>0$ (which implies $h>-1$),
spin up (PI3K) domain nucleation is started in the spin down (PTEN) sea.
The magnetization $m$ tends asymptotically to $h$, 
while the effective field $h_\mathrm{eff}$ tends to zero (Fig.~\ref{mag}), 
realizing the condition for phase coexistence.

\begin{figure}[htb]
\vspace{-0.3cm}
\includegraphics[height=18mm]{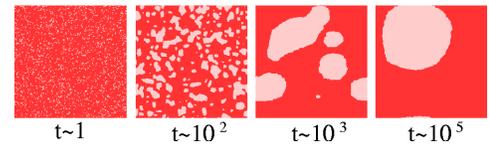}
\caption{%
Coarsening dynamics leading to random membrane polarization in the presence
of a uniform activation field.
}
\label{coars}
\end{figure}

After a rapid nucleation phase,
a domain coarsening dynamics follows: 
large domains grow and smaller ones shrink~\cite{LP81} (Fig.~\ref{coars}).

The final equilibrium state is characterized by 
the coexistence of the PI3K and the PTEN phase, localized in two
complementary clusters.
The equilibrium position of the 
PI3K cluster, which determines the direction of
cell movement, is random.
This behavior is consistent with experiments in which 
cells exposed to a uniform attractant distribution orient randomly
(stochastic polarization)~\cite{Zig77}.

\begin{figure}[htb]
\includegraphics[height=54mm]{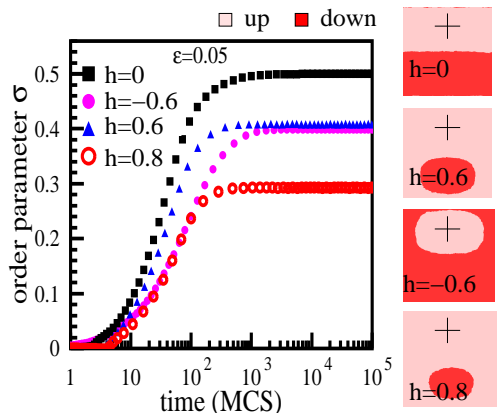}
\vspace{-0.3cm}
\caption{%
Time evolution of 
the order parameter for different values of the activation field $h$, 
and for a fixed value 
$\epsilon =0.05$ of the gradient.
At the end of the polarization process
the PI3K cluster (gray in the panels on the right) is centered around 
the point of maximum attractant stimulation (crosses).
}
\label{po}
\end{figure}

In the presence of a gradient in the chemical attractant
($\epsilon > 0$) the PI3K cluster localizes around the maximum of the 
attractant density.
To measure the polarization degree 
we define the following order parameter:
\begin{equation}
\sigma = \frac{1}{2}\frac{\sum_{i}^{N} (c_i-c) S_i}{\sum_{i}^{N} |c_i-c|},
\end{equation}
which is both a measure of
the degree of order in the system and 
of the correlation of the center of the 
PI3K cluster with the maximum of the attractant density (Fig.~\ref{po}).

\begin{figure}[htb]
\includegraphics[height=38mm]{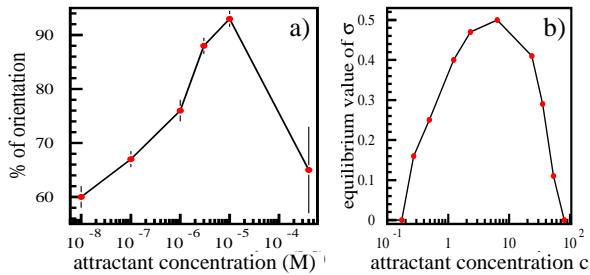}
\vspace{-0.3cm}
\caption{
a): Orientation degree of a population of cells 
as a function of the attractant concentration, for a constant gradient (adapted from~\cite{Zig77}).
b): Simulated equilibrium values of the order parameter
$\sigma$ as a function of the
attractant concentration $c$, for a constant gradient.
}
\label{doseh1}
\end{figure}

\textit{Dose-response curve --}
Simulations reproduce the qualitative behavior of experimentally
observed dose-response curves~\cite{SNB+06,Zig77}, showing no response for
either very high or very low values of the attractant concentration,
and optimal response for intermediate values (Fig.~\ref{doseh1}).
This effect can be explained as follows.
For very low $c$ the critical radius for patch nucleation is larger than 
the size of the cell, and no polarization is possible.
For very high $c$, such that $h>1$, the equilibrium magnetization is $1$, 
the whole system is uniformly populated by the PI3K phase, 
and again no polarization is possible.
Polarization is possible only for values which are intermediate between
these two limit cases.

\textit{Reversibility --}
Polarization induced by the gradient is reversible.
By changing the gradient direction after the system has reached
equilibrium, the position of the PI3K cluster adjusts to the new
direction in a finite time (not shown).
This effect reproduces the observed reorientation of eukaryotic cells
under varying attractant gradients observed in the experiments~\cite{JMD+04}.
Interestingly, after changing the sign
of the relative gradient 
we observed reorientation taking place by a collective movement of the
PI3K cluster, and not by its evaporation and successive recombination.

\begin{figure}[htb]
\includegraphics[height=45mm]{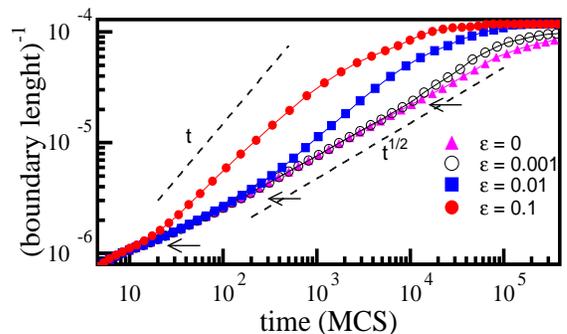}
\vspace{-0.5cm}
\caption{%
Time evolution of the inverse length of 
the total cluster boundary for different values of the gradient $\epsilon$.
The dotted lines show the slope of the 
power-law behaviors characterizing the growth regimes dominated,
respectively, 
by the uniform component of the attractant ($\sim t^{1/2}$) and
by the gradient ($\sim t$).
Arrows show the position of crossovers between the two scaling behaviors.
}
\label{feps}
\end{figure}

\textit{Gradient amplification and polarization time --}
The transient states are characterized by a coarsening
dynamics with the appearance
of scaling laws in the process of domain formation~\cite{LP81,GKL+07,Bra95}.
Our simulations
show that, for a condition of uniform distribution of attractant, in the 
initial coarsening stage 
the average cluster radius \(\langle r\rangle\) grows approximately as
\(t^{1/2}\) .
In Fig.~\ref{feps} the inverse length of the total cluster boundary is
plotted against time~\footnote{If the system is composed of circular domains, 
the inverse length of the total cluster boundary scales as the mean radius of the
clusters.}.

We define the polarization time $t_\mathrm{p}$ as the time for which the order parameter
$\sigma$ reaches 90\% of its equilibrium value.
If the attractant is uniformly distributed
the coarsening process stops
when the average cluster radius becomes
of the order of the cell size, 
\(r \sim L\), implying that the spontaneous cell polarization time scales as
\(t_\mathrm{p} \sim 1/L^2\).

\begin{figure}[htb]
\includegraphics[height=55mm]{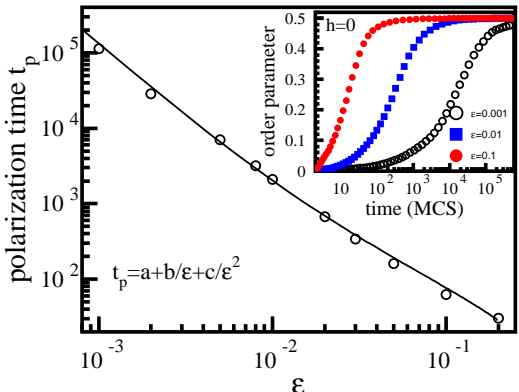}
\vspace{-0.4cm}
\caption{%
Polarization time 
as a function of $\epsilon$,
and time evolution of the order parameter $\sigma$ for
different values of $\epsilon$ (inset).
}
\label{potpol2}
\end{figure}

In the case of an attractant gradient we observe instead a double scaling behavior.
For $t<t_\epsilon$, where $t_\epsilon$ is a crossover time depending on the amplitude of the gradient $\epsilon$, cluster growth proceeds approximately as in the 
uniform case, while,
for $t>t_{\epsilon}$, 
the process of polarization becomes anisotropic, 
and the average cluster size grows approximately linearly in $t$ (Fig.~\ref{feps}).
The presence of this double scaling law implies that the polarization time 
behaves as $t_\mathrm{p}\sim a+b/\epsilon+c/\epsilon^2$
(Fig.~\ref{potpol2}).
We can understand the double scaling law as follows.
In the presence of an attractant gradient polarization takes place in two
steps.
In the initial (tuning) step the gradient of the attractant is negligible 
with respect to the uniform
component of the attractant and
cluster growth is approximately unaffected by its presence.
In the meanwhile, free enzymes shuttle from the cytosolic reservoir to
the membrane, lowering the chemical potential for further cluster growth
and effectively canceling out the effect of the uniform component of the
attractant.
This process continues until times of order $t_\epsilon$, when 
only the effect of the gradient component is
left. At this point, fast
polarization in the direction of the gradient takes place.

The anisotropic stage of cluster evolution 
leading to directed polarization
occurs only if 
$t_\epsilon<t_\mathrm{p}$. Otherwise, the presence of a gradient of attractant becomes
irrelevant and only the stage of isotropic patch growth actually occurs.
The crossover time $t_\epsilon$ increases with decreasing $\epsilon$
until it becomes of the order of $t_\mathrm{p}$,
implying the existence of a lower threshold
$\epsilon_\mathrm{th}$ of detectable gradients.
For \(\epsilon >\epsilon_\mathrm{th}\) 
anisotropy-induced polarization is 
much faster than spontaneous polarization. This 
explains the experimentally observed effect of gradient amplification in
chemotacting cells
and the observation~\cite{SNB+06} of a lower threshold of
detectable gradients, below which there is no directional sensing. 
Our results also confirm the theoretical predictions of~\cite{GKL+07}.

\textit{Discussion --}
We have introduced a 
simple lattice-gas
model for the process of 
eukaryotic
directional sensing,
which reproduces important aspects of the observed phenomenology 
and sheds light on 
the underlying physical mechanism.
The model maps signaling molecules and enzymes in spin variables, and
the effective interaction between enzymes on the membrane into
a ferromagnetic coupling.
Enzymes shuttling from the cytosolic reservoir to the membrane
is shown to provide a fundamental self-tuning mechanism which drives the system
towards phase coexistence and polarization, by counteracting the
effect of the external activation field.
In the presence of an attractant gradient this mechanism cancels out the
isotropic component of the attractant distribution in a first (tuning) stage of
cluster growth, preparing the ground for fast directed polarization in the
direction of the gradient in the next stage.
The control provided by enzyme shuttling is encoded in the coupling of the 
effective magnetic field
$h_\mathrm{eff}$ with the local order parameter $m$,
thus realizing an effective long-range repulsion between
enzymes and introducing in the model an element of self-organization.
The existence of two distinct stages in cluster evolution 
when an attractant gradient is present is signaled in the 
model simulations by the emergence of a double power
law for the time evolution of clusters of signaling molecules.
This shows up in the dependence of directed polarization time from the
gradient: for $\epsilon\ll 1$, $t_\epsilon$ scales as 
$\epsilon^{-2}$. 

Previous models of eukaryotic polarization
postulated the existence of a global inhibitor,
which was needed to cancel out 
the average value of 
the attractant leaving only the 
gradient component
(see~\cite{DJ03} for a review of previously proposed models of eukaryotic
chemotaxis).
The weak point of this approach is the necessity to fine-tune 
the activity of the global inhibitor in order to attain
perfect cancellation of the average attractant value.
In our scheme instead, the exchange
of chemical factors between the cell membrane and a finite cytosolic
reservoir realizes
a self-tuning mechanism which naturally leads to an equilibrium
state characterized by the coexistence of two distinct phases, similarly
to what happens in the case of first-order phase transitions in a
closed liquid-gas system or in the precipitation of a supersaturated solution.
The sensing mechanism encoded in the model is particularly robust,
allowing the cell to respond over a wide range of attractant concentrations.

Our model explains 
the experimentally observed behavior of chemotacting cells and reproduces
several effects, such as gradient amplification, typical 
dose-response curves, reversibility of orientation.
More generally, it shows that important biological functions 
may be described at a
physical level as self-organized phase transition processes.
 
\acknowledgments

We gladly acknowledge useful discussions with S.~Di~Talia,
I.~Kolokolov, V.~Lebedev and G.~Serini.


\vspace{-0.5cm}

\begin{thebibliography}{4}

\bibitem{RSB+03}A. Ridley \textit{et al.}, Science \textbf{302}, 5651 (2003).
\bibitem{LH96}D.A. Lauffenburger and A.F. Horwitz, Cell \textbf{84}, 359 (1996).
\bibitem{GCT+05}A. Gamba, \textit{et al.} Proc. Natl. Acad. Sci U.S.A.  \textbf{102}, 16927 (2005).
\bibitem{CGC+07}A. de Candia \textit{et al.}, Sci. STKE \textbf{378}, pl1 (2007). 
\bibitem{ID02}M. Iijima and P. Devreotes, Cell \textbf{109}, 599 (2002).
\bibitem{DMY+06}M. Dance \textit{et al.}, J. Biol. Chem. \textbf{281}, 23285 (2006).
\bibitem{Zig77}S.H. Zigmond, J. Cell. Biol. \textbf{75}, 606 (1977).
\bibitem{LP81}E.M. Lifshitz, and L.P. Pitaevskii, \textit{Physical Kinetics} (Pergamon Press, 1981).
\bibitem{SNB+06}L. Song \textit{et al.}, Eur. J. Cell Biol. \textbf{85}, 981 (2006).
\bibitem{JMD+04}C. Janetopoulos \textit{et al.}, Proc. Natl. Acad. Sci.  U.S.A. \textbf{101}, 16606 (2004).
\bibitem{Bra95}A. Bray, Adv. Phys. \textbf{45}, 357 (1994).
\bibitem{GKL+07}A. Gamba \textit{et al.}, Phys. Rev.  Lett. (2007).
\bibitem{DJ03}P. Devreotes and C. Janetopoulos, J. Biol. Chem. \textbf{278}, 20445 (2003).

%
%



%







\end{thebibliography}
\end{document}